\documentclass[aps,pra,twocolumn,showpacs,superscriptaddress,longbibliography]{revtex4-1}
\usepackage{graphicx}
\usepackage{adjustbox}
\usepackage{amsmath}
 \usepackage{booktabs}
 \usepackage{multirow}
\usepackage{amssymb}

\usepackage{epstopdf}

\usepackage{amsfonts}

\usepackage{dcolumn}

\usepackage{multirow}
\usepackage[utf8]{inputenc}
\usepackage{esint}
\usepackage{float}
\usepackage{breakurl}
\usepackage{hyperref}
\hypersetup{colorlinks,citecolor=blue}

\makeatletter
\@ifundefined{textcolor}{}
{
 \definecolor{BLACK}{gray}{0}
 \definecolor{WHITE}{gray}{1}
 \definecolor{RED}{rgb}{1,0,0}
 \definecolor{GREEN}{rgb}{0,1,0}
 \definecolor{BLUE}{rgb}{0,0,1}
 \definecolor{CYAN}{cmyk}{1,0,0,0}
 \definecolor{MAGENTA}{cmyk}{0,1,0,0}
 \definecolor{YELLOW}{cmyk}{0,0,1,0}
}

\usepackage{subfigure}\usepackage{epsfig}\usepackage{amsfonts}\usepackage{mathrsfs}
\usepackage[toc,page,title,titletoc,header]{appendix}
\usepackage{amsmath}

\makeatother

\begin{document}
\title{Accelerating Relaxation Dynamics in Open Quantum System with Liouvillian Skin Effect}
\author{Zeqing Wang}
\affiliation{Department of Physics, Renmin University of China, Beijing, 100872, China}
\affiliation{Shenzhen Key Laboratory of Ultraintense Laser and Advanced Material Technology, Center for Advanced Material Diagnostic Technology, and College of Engineering Physics, Shenzhen Technology University, Shenzhen 518118, China}

\author{Yao Lu} 
\affiliation{Shenzhen Institute for Quantum Science and Engineering (SIQSE), Southern University of Science and Technology, Shenzhen, P. R. China.}
\affiliation{International Quantum Academy, Shenzhen 518048, China.}
\affiliation{Guangdong Provincial Key Laboratory of Quantum Science and Engineering, Southern University of Science and Technology Shenzhen, 518055, China.}

\author{Yi Peng} 
\affiliation{Shenzhen Institute for Quantum Science and Engineering (SIQSE), Southern University of Science and Technology, Shenzhen, P. R. China.}
\affiliation{International Quantum Academy, Shenzhen 518048, China.}
\affiliation{Guangdong Provincial Key Laboratory of Quantum Science and Engineering, Southern University of Science and Technology Shenzhen, 518055, China.}

\author{Ran Qi}
\affiliation{Department of Physics, Renmin University of China, Beijing, 100872, China}

\author{Yucheng Wang} 
\email{wangyc3@sustech.edu.cn}
\affiliation{Shenzhen Institute for Quantum Science and Engineering (SIQSE), Southern University of Science and Technology, Shenzhen, P. R. China.}
\affiliation{International Quantum Academy, Shenzhen 518048, China.}
\affiliation{Guangdong Provincial Key Laboratory of Quantum Science and Engineering, Southern University of Science and Technology Shenzhen, 518055, China.}

\author{Jianwen Jie}
\email{Jianwen.Jie1990@gmail.com}
\affiliation{Shenzhen Key Laboratory of Ultraintense Laser and Advanced Material Technology, Center for Advanced Material Diagnostic Technology, and College of Engineering Physics, Shenzhen Technology University, Shenzhen 518118, China}
\date{\today}

\begin{abstract}
We investigate a non-Hermitian model featuring non-reciprocal gradient hoppings. 
Through an in-depth analysis of the Liouvillian spectrum and dynamics, we confirm the emergence of the Liouvillian skin effect resulting from the non-reciprocal nature of hoppings in this model. Furthermore, we observe that the presence of gradient hopping strength leads to an accelerated relaxation time for the system. Through numerical investigations of the {\color{black}Liouvillian gap}, relaxation time, and steady-state localization length, we discover that the relaxation time in this model cannot be explained by the currently established relationship associated with the Liouvillian skin effect. This discrepancy highlights the need for further exploration and theoretical advancements to fully comprehend the intricate mechanisms underlying quantum relaxation processes. Motivated by these findings, we propose a theoretical approach to realize this non-Hermitian model in an atomic system with a sideband structure by employing adiabatic elimination technique. These results contribute to our deeper comprehension of quantum relaxation dynamics and provide theoretical backing for the development of techniques aimed at controlling quantum relaxation processes.
\end{abstract}
\maketitle
\section{Introduction}
The study of open quantum systems, which takes into account the interactions with the surrounding environment, is a fundamental and captivating research field \cite{breuer2002theory,rivas2011open}. Many open quantum systems can be effectively described by the non-Hermitian Hamiltonians, which have attracted widespread attentions in the past two decades \cite{ashida2020non,zhang2022review,lin2023topological,PRL2016Lee,PRL2018Yao,PRL2019Song,PRB2022Peng}. Unlike closed quantum systems, open quantum systems experience a breakdown of time reversibility due to the stochastic coupling with the environment. This breakdown leads the open quantum system to eventually reach a steady state, where it remains throughout the evolution. This evolution is referred to as the relaxation process and occurs on a characteristic timescale known as the relaxation time, denoted as $\tau$. The relaxation time serves as a significant intrinsic timescale for understanding open quantum systems. In a specific class of open quantum systems characterized by the Markovian Lindblad master equation, the relaxation time $\tau$ is typically inversely proportional to the Liouvillian gap $\Delta$ of the system \cite{PRL2013Cai,PRE2015ifm}.


Furthermore, considerable attention has been given to the skin effect in Markov process-based open quantum systems. The existence of the skin effect in such systems has recently been confirmed and named the Liouvillian skin effect (LSE) \cite{PRL2021Ueda,PRR2022Yang}. Compared with the non-Hermitian skin effect (NHSE) that describes the localization of non-Hermitian Hamiltonian eigenstates \cite{ashida2020non,zhang2022review,lin2023topological,PRL2016Lee,PRL2018Yao,PRL2019Song,PRB2022Peng}, here the LSE denotes the localization of Liouvillian eigenmodes. In the LSE, the system tends to relax towards the boundaries of system. 
Interestingly, it has been discovered that the relaxation processes are slowed down in the presence of the LSE, even without the closing of the Liouvillian gap. The relationship between the relaxation time $\tau$ and the Liouvillian gap $\Delta$ is modified by the ratio of the system size $N$ to the localization length $\xi$ of the Liouvillian skin mode \cite{PRL2021Ueda}:
\begin{eqnarray}\label{rt}
\tau\sim\frac{1}{\Delta}(1+\frac{N}{\xi}).
\end{eqnarray}
This relationship significantly advances our understanding of relaxation physics in open quantum systems. It raises the question of its universality across all open quantum systems with the LSE. Moreover, if the relationship is not universal, it prompts further investigation into whether systems deviating from it exhibit even more intriguing Liouvillian dynamics.

\begin{figure*}[tbph]
\includegraphics[width=16cm]{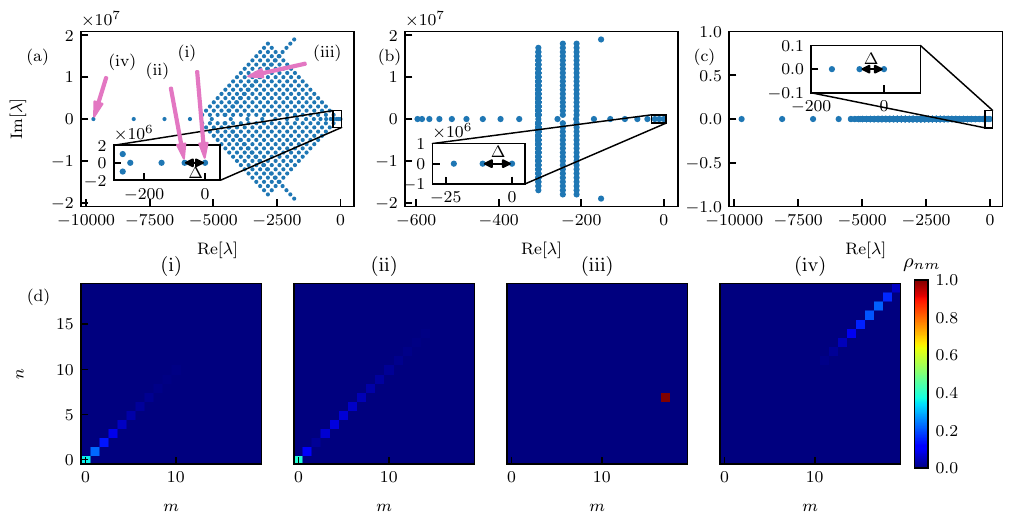}
\caption{(Color online) Liouvillian spectrum. The eigenvalues of Liouvillian operator: (a) with both on-site potential and nonreciprocal hoppings being gradient [$E_{n}=nE, J_{n,R(L)}=nJ_{R(L)}$]; (b) with only on-site potential being gradient [$E_{n}=nE, J_{n,R(L)}=J_{R(L)}$]; (c) with only nonreciprocal hoppings being gradient [$E_{n}=E, J_{n,R(L)}=nJ_{R(L)}$]. (d) The eigenmodes of Liouvillian operator as remarked in (a). The number of sites is $N=20$. Other parameters: $E=1.0$MHz, $J_{L}=184.3$Hz,  $J_{R}=118.0$Hz. }
\label{fig2}
\end{figure*}

To address these inquiries, we investigate a non-Hermitian model with non-reciprocal gradient hopping. Firstly, we establish the existence of the LSE by examining the Liouvillian eigenmodes and dynamics of the model. Additionally, we observe a significant acceleration of the relaxation process towards the steady state due to the presence of gradient non-Hermitian hopping, which modifies the relaxation relation stated in Eq. (\ref{rt}). Furthermore, we propose a method to implement this non-Hermitian model in atomic systems based on the sideband structure, utilizing the adiabatic elimination technique.

In the following, we introduce the non-Hermitian model and confirm the presence of the LSE through an analysis of the Liouvillian spectrum and dynamics in Section \ref{LSK}. We then delve into the investigation of the relaxation time of the non-Hermitian model in Section \ref{RT}. Next, we discuss our proposal for realizing the non-Hermitian model in atomic systems in Section \ref{model}. Finally, we present our concluding remarks in Section \ref{conclusion}.

\section{Liouvillian Skin effect}\label{LSK}
Here we study the non-Hermitian model described by the Lindblad master equation as following,
\begin{eqnarray}\label{ELMEx}
\dot{\hat{\rho}}=-i\left[\sum_{n=0}^{N-1} E_{n}|n\rangle\langle n|,\hat{\rho}\right]+ \sum_{n=1}\sum_{j=L,R}\mathcal{D}[\hat{L}_{n,j}]\hat{\rho},~~~
\end{eqnarray} 
where the Lindblad super-operator is defined as $\mathcal{D}[\mathcal{\hat{A}}]\hat{\rho} = \mathcal{\hat{A}}\hat{\rho} \mathcal{\hat{A}}^\dagger - \{{\mathcal{\hat{A}}^\dagger \mathcal{\hat{A}}, \hat{\rho} }/2\}$ and the Lindblad {\color{black}jump} operator $\hat{L}_{n,L(R)}$ denotes as
\begin{eqnarray}\label{ELOx}
\hat{L}_{n,L}=\sqrt{J_{n,L}}|n-1\rangle\langle n|,\hat{L}_{n,R}=\sqrt{J_{n,R}}|n\rangle\langle n-1|,~~~
\end{eqnarray} 
with the left hopping strength $J_{n,L}$ and right hopping strength $J_{n,R}$. $N$ is the number of site $\{|n\rangle\}$, i.e., the system size. 
In this model, both on-site energy $E_{n}$ and non-coherent hoppings $J_{n,R(L)}$ can be gradient. The Lindblad master equation we considering in Eq. (\ref{ELMEx}) can be rewritten as 
\begin{eqnarray}\label{ELME1}
\dot{\hat{\rho}}=\mathcal{L}[\hat{\rho}],
\end{eqnarray} 
where $\mathcal{L}$ is the Liouville super-operator defined in a $N^2$-dimensional Hilbert space \cite{breuer2002theory}. 
Then the right and left eigenmodes of $\mathcal{L}$ are defined as
\begin{eqnarray}\label{ELME2}
\mathcal{L}[\hat{\rho}^{r}_{k}]=\lambda_{k}\hat{\rho}^{r}_{k},\nonumber\\
\mathcal{L}^{\dagger}[\hat{\rho}^{l}_{k}]=\lambda_{k}^{*}\hat{\rho}^{l}_{k},
\end{eqnarray} 
with $k=0,1,2,\cdots,N^{2}-1$. Here $r(l)$ denotes to the normalized right (left) eigenmode as $\text{Tr}[\sqrt{(\hat{\rho}^{r(l)}_{k})^{\dagger}\hat{\rho}^{r(l)}_{k}}]=1$.  Thus, any initial state of the system $\hat{\rho}_{\text{ini}}$ can be expanded in terms of the eigenmodes as 
\begin{eqnarray}\label{ELME3}
\hat{\rho}_{\text{ini}}=\sum_{k=0}^{N^2-1}c_{k}\hat{\rho}^{r}_{k},
\end{eqnarray} 
where the coefficients $c_{k}$ are given by $c_{k}=\text{Tr}[(\hat{\rho}^{l}_{k})^{\dagger}\hat{\rho}_{\text{ini}}]/\text{Tr}[(\hat{\rho}^{l}_{k})^{\dagger}\hat{\rho}^{r}_{k}]$. As a result, the system evolves to the state
\begin{eqnarray}\label{ELME4}
\hat{\rho}(t)=\sum_{k=0}^{{\color{black}{N}}^2-1}c_{k}e^{\lambda_{k}t}\hat{\rho}^{r}_{k}.
\end{eqnarray} 
where $\lambda_k$ represents the decay rate associated with the eigenmode $\hat{\rho}^{r}_{k}$.

The time evolution of an open quantum system is characterized by quantum dynamical semigroups, which states that the fate of system is determined by the steady state $\hat{\rho}_{s}$, while the contributions from all other eigenmodes decay completely. The steady state $\hat{\rho}_{s}$ corresponds to the eigenmode of the Liouvillian superoperator $\mathcal{L}$ with a zero eigenvalue (excluding pure imaginary eigenvalues, which would lead to non-stationary steady states \cite{PRA2023Li,PRA2023Li2}). In other words, $\mathcal{L}[\hat{\rho}_{s}]=0$. This implies that the real parts of all other eigenvalues are negative, allowing us to order the eigenvalues $\lambda_{k}$ in descending order of their real parts as $0=\lambda_{0}>\text{Re}[\lambda_{1}]\ge\cdots\ge\text{Re}[\lambda_{N^{2}-1}]$. The Liouvillian gap, denoted as $\Delta=|\text{Re}[\lambda_{1}]|$, is defined as the real part of the eigenvalue of the Liouvillian superoperator with the largest nonzero real part. This gap is typically associated with the asymptotic decay rate \cite{PRA2018Min}. The time-dependent density matrix can be expressed as \cite{PRL2021Ueda}
\begin{eqnarray}\label{ELME5}
\hat{\rho}(t)=\hat{\rho}^{r}_{s}+\sum_{k=1}^{N^2-1}c_{k}e^{\lambda_{k}t}\hat{\rho}^{r}_{k}.
\end{eqnarray} 

To investigate the LSE in our system, we first numerically solve Eq. (\ref{ELME2}) to obtain the Liouville spectrum as shown in Fig. \ref{fig2}, considering a system with 20 sites. As depicted in Fig. \ref{fig2} (a), the eigenvalues with non-zero imaginary parts appear within a central region, indicating their contribution to the periodic oscillations in the relaxation dynamics. The inset plot confirms the uniqueness of the steady state, primarily resulting from the breaking of all the symmetries of system by the Lindblad {\color{black}jump} operator in Eq. (\ref{ELOx}) \cite{PRA2014Jiang}. In contrast, Fig. \ref{fig2} (b) displays the spectrum when the gradient of hoppings is turned off, resulting in reduced absolute values of the real parts of the eigenvalues. Furthermore, the distribution of the real parts of eigenvalue modes with non-zero imaginary parts in the middle section of the spectrum becomes more uniform, indicating a more consistent decay rate for these modes. Figure \ref{fig2} (c) demonstrates that when the on-site potential gradient is eliminated, eigenmodes with non-zero imaginary parts disappear.  In all cases shown in Figs. \ref{fig2} (a-c), the steady state remains unique, as depicted in the inset plot, indicating that the non-reciprocal hoppings do not alter the symmetry of system. Additionally, Figs. \ref{fig2} (a-b) show that the {\color{black}Liouvillian gap} is larger when gradient hopping is present, suggesting a faster relaxation rate towards the steady state. In Fig. \ref{fig2} (d), we present the density matrices of the eigenmodes labeled in Fig. \ref{fig2} (a). The steady state is localized at the left boundary of the system [see Fig. \ref{fig2} (d-i)], and as the real part of the eigenvalues increases, the corresponding eigenmodes tend to occupy sites near the right boundary [see Fig.1(d-iv)]. For eigenmodes with non-zero imaginary parts of eigenvalues, their density matrices exhibit non-zero off-diagonal elements, leading to oscillatory decaying behavior during the relaxation process towards the steady state.
\begin{figure}[tbph]
\includegraphics[width=8.6cm]{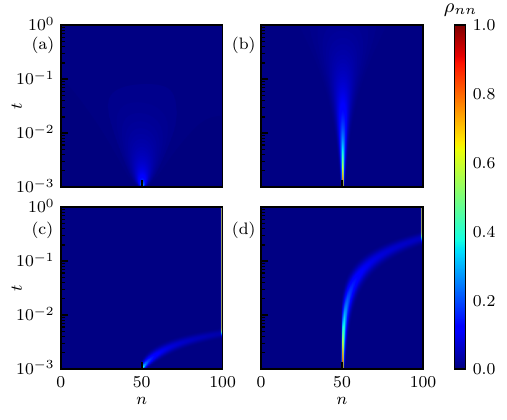}
\caption{(Color online) Liouville dynamics. The hopping strengths are gradient in (a,c) [$J_{n,R(L)}=nJ_{R(L)}$] and are homogeneous in (b,d) [$J_{n,R(L)}=J_{R(L)}$]. (a,b) show the absence of LSE with hopping strengths $J_{R}=J_{L}=184.3$Hz, and (c,d) confirm the LSE with nonreciprocal hopping strengths $J_{R}=100J_{L}=184.3$Hz. The number of sites is $N=100$. Other parameters: $E_{n}=nE,~E=1.0$MHz. The panels in the same row share the same $y$-axis.}
\label{fig3}
\end{figure}

Taking the system size as $N=101$ and setting the initial state of the system to $|n=50\rangle$, we make noteworthy observations. When the system undergoes reciprocal hoppings [Figs. \ref{fig3}(a-b)], it displays symmetric dynamical evolution across the system. However, in the case of non-reciprocal hoppings [Figs. \ref{fig3}(c-d)], the system's symmetric dynamical evolution breaks down, and it evolves towards the boundaries, remaining there indefinitely. This intriguing phenomenon is known as the LSE. Furthermore, consistent with the findings in Fig. \ref{fig2}, when the hoppings in the system are gradient [Fig. \ref{fig3}(a,c)], the system relaxes to the boundaries at a faster rate. As the hoppings in our model exhibit a gradient nature, we will quantitatively study the key factors influencing the relaxation process under such special hoppings: the {\color{black}Liouvillian gap}, relaxation time, and localization length.
\section{Relaxation time}\label{RT}
\begin{figure}[tbph]
\includegraphics[width=8.6cm]{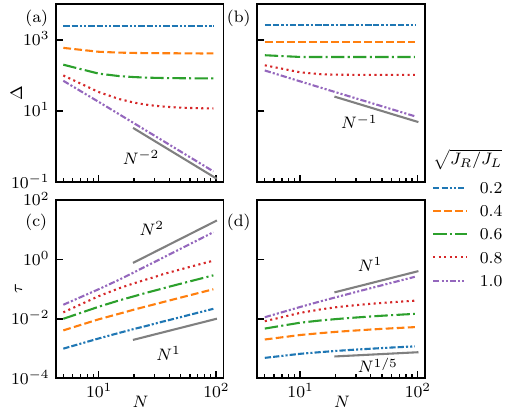}
\caption{(Color online) Liouvillian gap and relaxation time. The hopping strengths are homogeneous in (a,c) [$J_{n,R(L)}=J_{R(L)}$] and are gradient in (b,d) [$J_{n,R(L)}=nJ_{R(L)}$]. 
The solid gray lines give the reference of the size scaling. Other parameters: $E_{n}=nE,~E=1.0$MHz, $J_{L}=184.3$Hz. The panels in the same row share the same $y$-axis.}
\label{fig4}
\end{figure}
The relaxation time in a quantum open system refers to the characteristic duration it takes for the system to reach its equilibrium or steady state \cite{breuer2002theory}. This time span is influenced by several factors, including the strength of the system's interaction with the environment, the properties of the environment itself, and the specific dynamics governing the system. Experimental determination of the relaxation time involves observing the temporal evolution of relevant observables or analyzing the decay rates of specific quantities. Understanding the relaxation time is crucial in the study of open quantum systems as it provides valuable insights into system behavior, stability, and the timescales associated with achieving a steady state. Moreover, it holds significant importance in practical applications like quantum information processing, where effective control and mitigation of relaxation processes are essential for preserving the coherence and reliability of quantum states and operations \cite{Harrington2022}.

To discuss the variation of the {\color{black}Liouvillian gap} and relaxation time with system size and the hopping strength, we consider, without loss of generality,  the case that the steady state of the system $\rho_{s}$ localizes on the left boundary site $|0\rangle$, namely  $J_{n,L} > J_{n,R}$. Therefore, we initialize the system in the right boundary site $|N-1\rangle$ and define the relaxation time $\tau$ as the time when the decay of the population of the right boundary site $|N-1\rangle$ reaches $1/e$ of $\rho_{s,N-1}$.  Then the the relaxation time $\tau$ is given by 
 \begin{eqnarray}\label{ELME6}
\rho_{s,N-1}-\rho_{N-1}(\tau)=\frac{\rho_{s,N-1}}{e},
\end{eqnarray} 
where $\rho_{s,n}$ represents the occupation probability of the steady state on the site $|n\rangle$.
\begin{figure}[tbph]
\includegraphics[width=8.6cm]{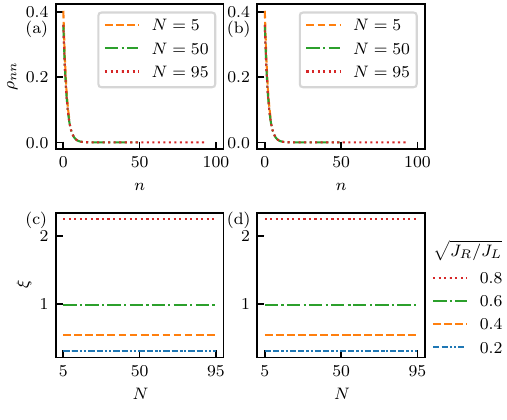}
\caption{(Color online) Localization length. The hopping strengths are homogeneous in (a,c)  [$J_{n,R(L)}=J_{R(L)}$] and are gradient in (b,d)  [$J_{n,R(L)}=nJ_{R(L)}$].  (a-b) show the profile of steady states for different system size with $\sqrt{J_{R}/J_{L}}=0.8$. (c-d) show the localization length of the steady states.
 Other parameters are same to Fig. \ref{fig4}. The panels in the same row share the same $y$-axis.}
\label{fig5}
\end{figure}

In Fig. \ref{fig4}, we consider two cases: homogeneous hopping strength (a, c) and gradient hopping strength (b, d). When the system undergoes reciprocal and homogeneous hoppings [double-dotted dashed purple lines in Figs. \ref{fig4} (a) and (c)], we observe that the relaxation time $\tau$ (proportional to $N^{2}$) and the {\color{black}Liouvillian gap} $\Delta$ (proportional to $N^{-2}$) follow the relationship $\tau \propto 1/\Delta$, reflecting {\color{black}diffusive relaxation \cite{Agmon1985,spiechowicz2016transient,PRL2021Ueda,PRE2015S}}. However, when the hopping strength is gradient and reciprocal, the system relaxes to the steady state at an accelerated rate ($\tau \propto N$), as indicated by the double-dotted dashed purple lines in Figs. \ref{fig4} (b, d). Interestingly, when the hopping strength is non-reciprocal and homogeneous, the simple relationship $\tau \propto 1/\Delta$ is broken. As non-reciprocity increases, the value of $\Delta$ tends to remain invariant with the system size [see the double-dotted double-dashed blue lines in Figs. \ref{fig4} (a-b)], while the relaxation time $\tau$ still scales with the system size [see the double-dotted double-dashed blue lines in Figs. \ref{fig4} (c-d)]. In this case, the relationship between the {\color{black}Liouvillian gap} and the relaxation time needs to be described by Eq. (\ref{rt}), where the localization length $\xi$ is size-independent. 
However, as shown in Fig. \ref{fig4}(b) and Fig. \ref{fig4}(d), when the hopping is gradient, with increasing non-reciprocity, the value of $\Delta$ still does not change with the system size, but the relaxation time $\tau$ scales as $N^{1/5}$. This result indicates that the gradient significantly accelerates the relaxation process. Furthermore, if the system still follows Eq. (\ref{rt}), i.e., $\tau\sim\Delta(1+N/\xi)\propto N^{1/5}$, it implies that the localization length of the system $\xi$ is size-dependent and scales as $\xi\sim N^{4/5}$.

To verify whether the localization length of the system follows the above analysis, we present the plot of the localization length as a function of size in Figs. \ref{fig5}(a-b). We observe that when the system exhibits the LSE, regardless of whether the hopping strength is homogeneous [Fig. \ref{fig5}(a)] or gradient [Fig. \ref{fig5}(b)], the profile of steady states for smaller systems is overlapped by the profile of steady states for larger systems, indicating that the localization length of the steady state is independent of the system size. This size-independent behavior is further supported by Figs. \ref{fig5}(c-d), which show that the localization length of the steady state is solely determined by the nonreciprocal hopping ratio $J_{R}/J_{L}$, regardless of whether the hopping strength is homogeneous [Fig. \ref{fig5}(c)] or gradient [Fig. \ref{fig5}(d)]. Therefore, the relaxation behavior of our model cannot be described by Eq. (\ref{rt}). This indicates that the general relationship governing the relaxation time for open quantum systems with the LSE has not yet been discovered.

\begin{figure}[tbph]
\includegraphics[width=8.6cm]{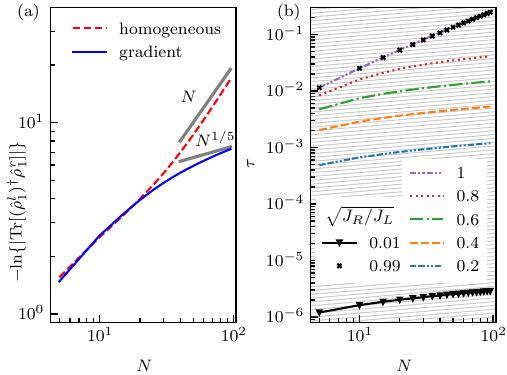}
\caption{{\color{black}(Color online) (a) Overlap of the left eigenmode and the right eigenmode for different system size with $\sqrt{J_{R}/J_{L}}=0.8$.(b) Relaxation time for different system size.  The solid gray lines give the reference of the size scaling $N^{1/5}$.  Other parameters are same to Fig. \ref{fig4}.}}
\label{fig66}
\end{figure}

{{\color{black}To gain further insight into the physics underlying our results, we closely follow the analysis presented in \cite{PRL2021Ueda}. The acceleration of the relaxation process can be understood using Eq. (\ref{ELME5}). Among all the eigenmodes, the real part of the eigenvalues of $\rho_1^{r}$ has the smallest absolute value, leading to the slowest relaxation to the steady state. Consequently, the entire relaxation time scale is determined by the coefficient corresponding to $\rho_1^{r}$, denoted as $c_1$. It is always possible to prepare the initial state $\hat{\rho}_{\text{ini}}$ with an overlap of $O(1)$ with $\rho_1^{l}$. When the system exhibits Liouvillian skin effect (LSE) under homogenous hopping, all the eigenmodes are localized near the boundary and decay exponentially, resulting in $c_1 =O(1)/\text{Tr}[(\hat{\rho}^{l}_{1})^{\dagger}\hat{\rho}^{r}_{1}]\sim e^{O(N/\xi)}$ (without LSE, $c_1$ is independent of the system size $N$, and the relaxation law is $1/\Delta$). At this point, the relaxation time $\tau$ is determined by $e^{O(N/\xi)}e^{-\tau\Delta}\sim e^{-1}$, leading to the relation given in Eq. (\ref{rt}). In Fig. \ref{fig66}(a), we numerically calculate $-\ln{|\text{Tr}[(\hat{\rho}^{l}_{1})^{\dagger}\hat{\rho}^{r}_{1}]|}$ and verify the size scaling $N$ for the homogenous hopping. Furthermore, we observe that for the gradient hopping, $-\ln{|\text{Tr}[(\hat{\rho}^{l}_{1})^{\dagger}\hat{\rho}^{r}_{1}]|}$ shows a tendency to converge to $N^{1/5}$ as the system size increases. Following the analysis above, we infer that $c_1 =O(1)/\text{Tr}[(\hat{\rho}^{l}_{1})^{\dagger}\hat{\rho}^{r}_{1}]\sim e^{O(N^{1/5}/\xi)}$ for the gradient hopping. Consequently, based on the results in Figs. (\ref{fig4}-\ref{fig66}), we deduce the relaxation time for our model as $\tau\sim\Delta^{-1}(1+N^{1/5}/\xi)$ in the thermalization limit. We numerically check the size scaling $N^{1/5}$ in Fig. \ref{fig66}(b), where the relaxation time shows the tends of convergence to the $N^{1/5}$ in the larger system size. This can be understood as the system size $N$ being effectively shortened to $N^{1/5}$ by the gradient hopping, which breaks the translation symmetry of the system by inducing an additional effective force. As a result, regardless of whether the system evolves from the left to the right or from the right to the left, the overall dynamics are accelerated, with the only difference being the time stage for the accelerations. To precisely verify and prove our results, we need to find new methods to analytically solve Eq. (\ref{ELMEx}). This remains our primary focus for future research.}

\begin{figure*}[tbph]
\includegraphics[width=17cm]{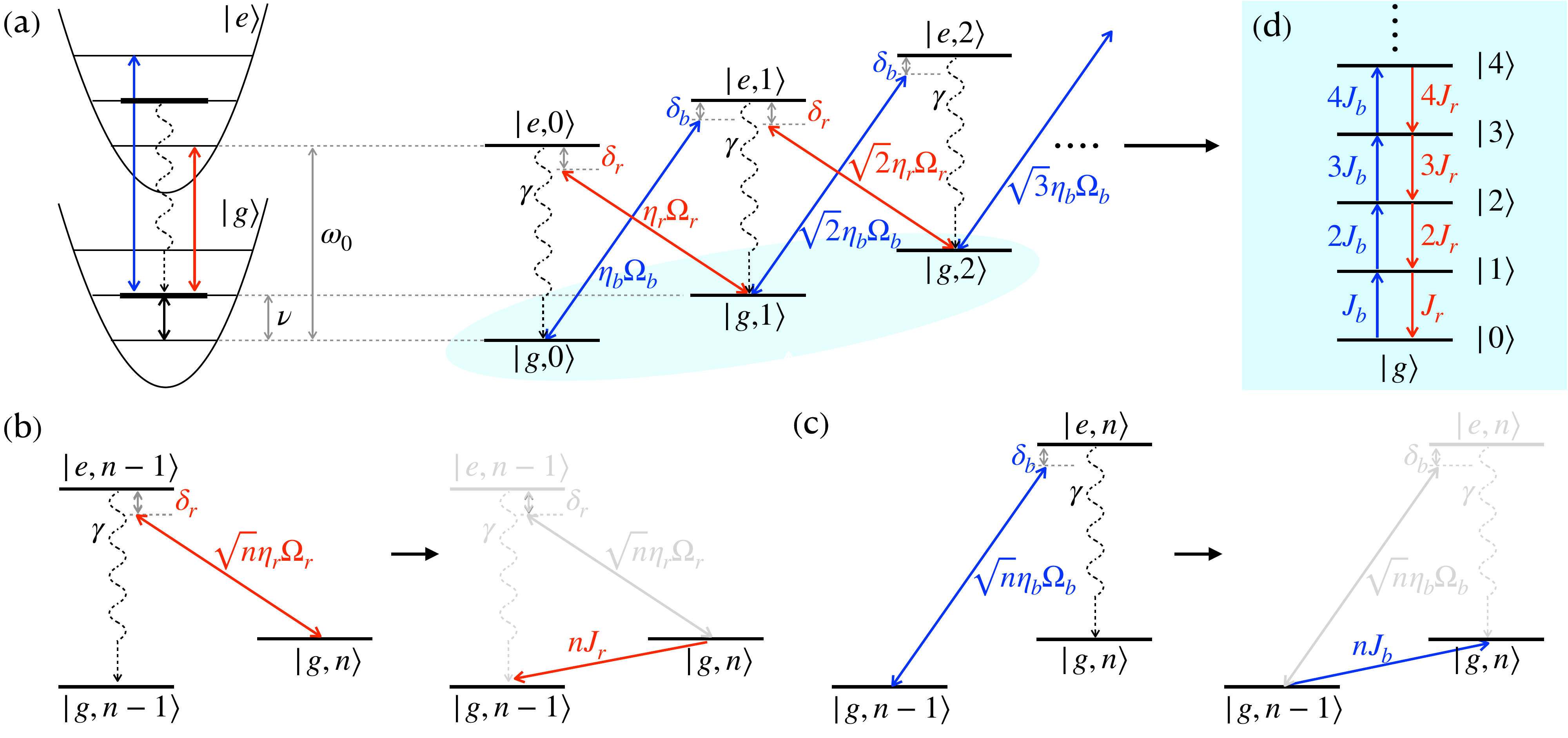}
\caption{(Color online) Illustration of the proposed non-Hermitian model. (a) The motional sidebands structure of a trapped-ion system encoded with two internal states $|g\rangle$ and $|e\rangle$. $\Delta$ and $\nu$ are the level spacings of the internal states and the motional sidebands states, respectively. $\delta_{r,b}$, $\eta_{r,b}$, and $\Omega_{r,b}$ are respectively the detunings, Lamb-Dick parameters, and rabi frequencies of the independent red-detuned laser and blue-detuned laser. $\gamma$ is the decay rate of the excited internal state $|e\rangle$. 
(b) and (c) show the adiabatical elimination processes of the excited state $|e,n-1\rangle$ via red-detuned laser and of the excited state $|e,n\rangle$ via blue-detuned laser. $nJ_{r,b}$ is the effective hopping strength. (d) The ground internal state $|g\rangle$ based effective non-Hermitian model with non-reciprocal gradient hoppings $nJ_{r,b}$.}
\label{fig1}
\end{figure*}
\section{Proposal of non-Hermitian model with gradient hopping}\label{model}
We use the trapped-ion system as an example to illustrate the effective non-Hermitian model. The method employed here is applicable to other atomic systems that possess the motional sideband structure \cite{PRL1998Ha,PRX2012Kau,PRA2022Zo}. As depicted in Fig. \ref{fig1}(a), the trapped-ion system consists of two internal electronic energy levels: the ground state $|g\rangle$ and the excited state $|e\rangle$, which are described by the Hamiltonian (we set $\hbar=1$ throughout this work),
\begin{eqnarray}
\hat{H}_{i}=\frac{\omega_{0}}{2}\left(|e\rangle\langle e|-|g\rangle\langle g|\right).
\end{eqnarray}   
The trap employed here provides dynamical confinement in the $y$-$z$ plane and static confinement in the $x$ direction \cite{RMP2003Lei}. The motional sidebands of the internal states are constructed using the energy levels $\{|n\rangle, n=0,1,2,\cdots\}$ of the harmonic trap in the $x$ direction, which are separated by the frequency $\nu$ and given by
\begin{eqnarray}
\hat{H}_{e}=\sum_{n=0}^{N-1}\left(\frac{1}{2}+n\right)\nu|n\rangle\langle n|,
\end{eqnarray}   
with the system size $N$ (the number of motional sideband levels). We introduce two independent lasers for coupling the internal states to the external motional sideband states. The couplings are described by
\begin{eqnarray}
\hat{V}_{j}=\Omega_{j}\left(|g\rangle\langle e|+|e\rangle\langle g|\right)\cos\left(k_{j}\hat{x}_{S}-\omega_{j}t+\phi_{j}\right),
\end{eqnarray}  
where $k_{j}, \omega_{j}, \phi_{j}$, and $\Omega_{j}$ correspond to the wave vector, frequency, initial phase, and Rabi frequency of laser $j$, respectively. Here, the subscripts $j=r,b$ refer to the red-detuned laser ($\omega_{r}<\omega_{0}$) and the blue-detuned laser ($\omega_{b}>\omega_{0}$), respectively. 

We apply the rotating wave approximation to the system in the rotating frame $\hat{U}_{R}=e^{-i(H_{i}+H_{e})t}$, resulting in the Hamiltonian of the trapped-ion:
\begin{eqnarray}\label{HRWA}
\hat{H}_{R}^{\text{RWA}}=\sum_{j=r,b}\frac{\Omega_{j}}{2}e^{-i\left[k_{j}\hat{x}_{R}-(\omega_{j}-\omega_{0})t+\phi_{j}\right]}|g\rangle\langle e|+\text{h.c.},
\end{eqnarray} 
with $\hat{x}_{R}=\hat{U}_{R}^{\dag}\hat{x}_{S}\hat{U}_{R}$. The system enters the Lamb-Dicke regime when the spatial extension of the ion $x_{0}=\sqrt{1/2M\nu}$ ($M$ is the mass of the ion) is much smaller compared to the wavelengths of all the applied lasers. In this regime, the recoil energies of the lasers have a negligible impact on the trap frequency $\nu$. Hence, the Lamb-Dicke parameter satisfies $\eta_{j}=k_{j}x_{0}\ll1~(j=r,b)$. We can then expand Eq. (\ref{HRWA}) in terms of $\eta_{j}$ to obtain the Hamiltonian:
\begin{eqnarray}\label{HRLD}
&&\hat{H}_{R}^{\text{LD}}=\sum_{n=0}\frac{\sqrt{n + 1}}{2}\left[\eta_{r}\Omega_{r}e^{i(\delta_{r}t-\phi_{r})}|g,n+1\rangle\langle e,n|+\text{h.c.}\right]\nonumber\\
&&+\sum_{n=0}\frac{\sqrt{n+1}}{2}\left[\eta_{b}\Omega_{b}e^{i(\delta_{b}t-\phi_{b})}|g,n\rangle\langle e,n+1|+\text{h.c.}\right],
\end{eqnarray} 
where $\delta_{r}=\omega_{r}-(\omega_{0}-\nu)$ and $\delta_{b}=\omega_{b}-(\omega_{0}+\nu)$ are respective the red detuning and blue detuning of lasers, and satisfy $\delta_{r,b}\ll\nu$.

The spontaneous decay of the excited sideband state $|e,n\rangle$ to the ground sideband state $|g,n\rangle$ with a decay rate $\gamma$ can be described by the Lindblad operators:
\begin{eqnarray}\label{Ln}
\hat{L}_{n}=\sqrt{\gamma}|g,n\rangle\langle e,n|.
\end{eqnarray} 
This leads to the Lindblad master equation for the system:
\begin{eqnarray}\label{LME}
\dot{\hat{\rho}}_t=-i\left[\hat{H}_{R}^{\text{LD}},\hat{\rho}_t\right]+ \sum_{n=0} \mathcal{D}[\hat{L}_n]\hat{\rho}_t,
\end{eqnarray} 
where the Lindblad super-operator $\mathcal{D}[\mathcal{\hat{A}}]\hat{\rho}_t = \mathcal{\hat{A}}\hat{\rho}_t \mathcal{\hat{A}}^\dagger - \{{\mathcal{\hat{A}}^\dagger \mathcal{\hat{A}}, \hat{\rho}_t }/2\}$.

Our discussions are based on the resolved motional sidebands, which requires the system both working in the Lamb-Dicke regime and satisfying the energy scale relations $\gamma\ll\nu$. Furthermore, we are interested in the weak motional sidebands coupling regime given by $\eta_{j}\Omega_{j}\ll\gamma$ \cite{Zhang_2021}. Therefore, as shown in Fig. \ref{fig1}(b), the trapped-ion will immediately decay to $|g,n-1\rangle$ following the red bandside hopping from $|g,n\rangle$ to $|e,n-1\rangle$. Exploiting this fact, we can adiabatically eliminate the unstable excited sideband states $|e,n-1\rangle$ and obtain the effective unidirectional hopping from $|g,n\rangle$ to $|g,n-1\rangle$ with strength $nJ_{r}$. These processes form a dissipative cascade, cooling the system to the ground state $|g,0\rangle$. Figure \ref{fig1}(c) illustrates similar processes for the blue bandside hoppings, which results in effective unidirectional hopping from $|g,n-1\rangle$ to $|g,n\rangle$ with strength $nJ_{b}$ and construct a gain cascade, heating the system to ground sideband states with higher energy. Then, as depicted in Fig. \ref{fig1}(d), the effective non-Hermitian model of the ground sideband state $|g,n\rangle$ comprises a semi-infinite ladder ${|n\rangle}$ with non-reciprocal blue-detuned gradient hopping ${nJ_{b}}$ and red-detuned gradient hopping ${nJ_{r}}$. According to the spirit of the adiabatic elimination method \cite{PRA2012Florentin}, the effective master equation in the Schr$\ddot{\text{o}}$dinger picture for this model can be written as (see Appendix \ref{App1} for details):
\begin{eqnarray}\label{ELME}
\dot{\hat{\rho}}=-i\left[\hat{H}_{\text{eff}},\hat{\rho}\right]+ \sum_{n=1}\left(\mathcal{D}[\hat{L}_{n,r}]\hat{\rho}+\color{black}\mathcal{D}[\hat{L}_{n,b}]\hat{\rho}\right),
\end{eqnarray} 
where $\hat{L}_{n,r(b)}$ is the effective Lindblad operator and denotes as
\begin{eqnarray}\label{ELO}
\hat{L}_{n,r}=\sqrt{nJ_{r}}|n-1\rangle\langle n|,~~\hat{L}_{n,b}=\sqrt{nJ_{b}}|n\rangle\langle n-1|,~~
\end{eqnarray} 
with the effective hopping strengths 
\begin{eqnarray}\label{JrJb}
J_{r(b)}=\frac{\gamma|\Omega_{r(b)}|^{2}\eta_{r(b)}^{2}}{4\delta_{r(b)}^{2}+\gamma^{2}}.
\end{eqnarray} 
The effective Hamiltonian reads (here we have removed the constant terms)
\begin{eqnarray}\label{KJAS}
\hat{H}_{\mathrm{eff}}=\hat{U}_R \hat{H}_{R}^{\text{LD}} \hat{U}_{R}^{\dagger}
=\sum_{n=0} n(E_r + E_{b} + \nu) |n\rangle\langle n|,
\end{eqnarray}  
where $E_{r(b)}=\delta_{r(b)}|\Omega_{r(b)}|^{2}\eta_{r(b)}^{2}/(4\delta_{r(b)}^{2}+\gamma^{2})$ is the energy shift induced by the lasers. 

We have derived a non-Hermitian model described by Eq. (\ref{ELME}) in the trapped-ion systems. In this model, the hoppings between different sideband levels are governed by a non-Hermitian quantum {\color{black}jump} operator, and the non-reciprocal hopping strength is achieved by adjusting parameters such as the Rabi frequency of the laser. This non-Hermitian model is useful for describing the sideband cooling and related sideband phonon excitation effects \cite{scully1997quantum, RMP2003Lei,Zhang_2021,PRL1995Monroe,PR2008H}. 

\section{Conclusion}\label{conclusion}
We have investigated a non-Hermitian model featuring non-reciprocal gradient hoppings, revealing the presence of the LSE through analysis of the Liouvillian spectrum and dynamics \cite{PRL2021Ueda}. Furthermore, we have observed that the gradient hopping strength in this model leads to an accelerated relaxation time of the system. However, our numerical investigations of the {\color{black}Liouvillian gap}, relaxation time, and steady-state localization length have shown that the currently known relaxation relation associated with the LSE does not fully explain the behavior observed in this model. These findings deepen our understanding of quantum relaxation dynamics and provide theoretical support for the development of techniques aimed at controlling and manipulating quantum relaxation processes. We have also proposed a theoretical scheme, based on the sideband structure, to implement this non-Hermitian model by using adiabatic elimination method. While we have illustrated the proposal using the trapped-ion system as an example, the method is applicable to other atomic systems possessing the motional sideband structure. Future research directions include gaining a more precise understanding of the observed acceleration phenomenon, deriving a general expression for the relaxation time, and exploring relaxation dynamics in other non-Hermitian models.
\begin{acknowledgments}
This work was supported by the National Natural Science Foundation of China (Grant No. 12104210, Grant No. 12104205, Grant No. 12022405,), 
the National Key R and D Program of China (Grant No. 2018YFA0306502, Grant No. 2022YFA1405800), the Shenzhen Science and Technology Program under (Grant No. ZDSYS20200811143600001, {\color{black}Grant No. RCBS20200714114820298}).
\end{acknowledgments}

\global\long\def\id{\mathbbm{1}}
\global\long\def\ui{\mathbbm{i}}
\global\long\def\ud{\mathrm{d}}

\setcounter{equation}{0}
\setcounter{figure}{0}
\setcounter{section}{0}
\setcounter{table}{0} 
\renewcommand{\theparagraph}{\bf}
\renewcommand{\thefigure}{S\arabic{figure}}
\renewcommand{\theequation}{S\arabic{equation}}

\onecolumngrid
\flushbottom
\appendix
\section{Derivation of the effective master
equation in Eq. (\ref{ELME})}\label{App1}
\twocolumngrid
In this section, we will derive the effective master equation as shown in
 Eq.\eqref{ELME} in the main text by the effective operator formalism for open
quantum systems \cite{PRA2012Florentin}.

Our system consist of two distinct subspaces, i.e. $|g, n\rangle$ and $|e, n\rangle$.
In the rotating frame $\hat{U}_R$, the Hamiltonian only contins the perturbative
coupling between these two subspaces. We firstly rewrite Eq.\eqref{ELME} as
\begin{align}
  \hat{H}_{R}^{\text{LD}} =
  \sum_{j=r, b}\sum_{n=0}
  \hat{V}_{+}^{(n, j)}(t)  + \mathrm{h.c.},
\end{align}
where $\hat{V}_{+}^{(n, j)}(t) = \hat{v}_+^{(n, j)} e^{-i\delta_j t}$ is a time
dependent perturbative field applied to couple the $|g, n\rangle$ to $|e, n\rangle$. Each oscillator state $|n\rangle$ is coupled by two laser fields, i.e. a red-detuned laser and a blude detuned laser, being labeled as $j= r, b$,
\begin{align}
\hat{v}_{+, r}^{(n)} =& \sqrt{n+1}\frac{\Omega_r}{2}\eta_r e^{i \phi_r} |e,
                        n\rangle\langle g, n+1|, \\
  \hat{v}_{+, b}^{(n)} =& \sqrt{n+1}\frac{\Omega_b}{2}\eta_b e^{i \phi_b} |e,
  n+1\rangle\langle g, n|.
\end{align}
Here we consider the Lindblad operators in the rotating frame $\hat{U}_R$ as
\begin{align}
\hat{L}_{n, R} = \sqrt{\gamma} e^{- i\omega_0 t} |g, n\rangle\langle e, n|.
\end{align}
Then we can perform the adiabatic elimination to arrive at an effective master equation
for the subspace $\{|g, n\rangle\}$ as
\begin{eqnarray}\label{eqS5}
  \dot{\hat{\rho}}_R=-i\left[\hat{H}_{R, \text{eff}},\hat{\rho}_R\right]+
  \sum_{n=1}\mathcal{D}[\hat{L}_{n,R, \mathrm{eff}}]\hat{\rho}_{R},
\end{eqnarray}
where the effective Hamiltonian in the rotating frame $\hat{U}_R$ is given by 
\begin{eqnarray}
  \hat{H}_{R, \mathrm{eff}} = -\frac{1}{2} \left[
  \hat{V}_-(t) \sum_{j=r, b}\sum_{n=0} (\hat{H}_{\mathrm{NH}}^{(j)})^{-1}
   \hat{V}_+^{(n, j)}(t) +\mathrm{h.c.}
  \right]~~~~
\end{eqnarray}
with the transition operator $\hat{V}_-(t) = \sum_{j=r, b}\sum_{n=0} \hat{V}_-^{(n, j)}(t)$, which describes the the effective transition process from $|g, n\rangle$ to $|e, n\rangle$ and then back to $|g, n\rangle$. The strength of this effective transition is determined by the propagator
\begin{align}
  (\hat{H}_{\mathrm{NH}}^{(j)})^{-1} =&
   \sum_{m=0}|e, m\rangle \frac{1}{-
  \frac{i}{2}\hat{L}^{\dagger}_m \hat{L}_m- \delta_j}\langle e, m|,\nonumber \\
 =&  \sum_{m=0}|e, m\rangle \frac{1}{-
  \frac{i}{2}\gamma - \delta_j}\langle e, m|.
\end{align}
Then we can straightfowardly obtain the following effective Hamiltonian
\begin{eqnarray}\label{SHeff}
  \hat{H}_{R, \mathrm{eff}} &&=\sum_{n=0} \left[  nE_{r} + (n+1)E_{b} \right] |g, n\rangle\langle g, n|\nonumber \\
     &&+ \sum_{n=0} \left[J_{n, n + 2} e^{-\mathrm{i}(\delta_b - \delta_r)t}|g, n + 2 \rangle\langle g, n|+ \mathrm{h.c.}\right],~~~~~
\end{eqnarray}
where
\begin{align}
  E_{r(b)} =& |\Omega_{r(b)}|^2 \eta_{r(b)}^{2}\frac{\delta_{r(b)}}{4\delta_{r(b)}^2 + \gamma^2}, \\
  J_{n, n+2} =& e^{\mathrm{i}(\phi_b - \phi_{r})}\sqrt{(n + 1)(n + 2)} \Omega_{r}\Omega_{b} \eta_r\eta_b\nonumber \\
     &\times \frac{(\delta_r + \delta_b)/2}{4\delta_r\delta_b + \gamma^2 + 2\mathrm{i}\gamma(\delta_{r} - \delta_b )}.
\end{align}
The first term in Eq. (\ref{SHeff}) is the on-site energy shift induced by the two lasers. The second term is corresponding to the long-range coherent hoppings comprising one blue detuning transition process and one red detuning transition process. 

The dissipative part of Eq. (\ref{eqS5}) is determined by the effective Lindblad operators in the rotating frame $\hat{U}_R$, which are given by 
\begin{align}\label{erS11}
  \hat{L}_{n, R, \mathrm{eff}} =&&\hat{L}_{n, R} \sum_{j=r, b}\sum_{m=0}
  (\hat{H}_{\mathrm{NH}}^{(j)})^{-1} \hat{V}_+^{(m, j)}(t),\nonumber\\
 =&& \sqrt{(n + 1) }j_r e^{-i(\delta_r + \omega_0) t}|g, n \rangle\langle g, n + 1|\nonumber\\
&& + \sqrt{n} j_be^{-i(\delta_b + \omega_0) t}|g, n\rangle\langle g, n - 1|,
\end{align}
where
\begin{align}
  j_{r(b)}
  =& \frac{\sqrt{\gamma}}{- i\gamma/2 - \delta_{r(b)}}\frac{\eta_{r(b)} \Omega_{r(b)}e^{i\phi_{r(b)}}}{2}.
\end{align}
We will obtain the time-independent terms proptional to $|j_{b(r)}|^2$ and the time-dependent cross terms ($\sim e^{-i(\delta_{r}-\delta_{b})t}$) of $j_{r}$ and $j_{b}$, when we expand the Lindblad super-operator $\mathcal{D}[\hat{L}_{n,R,\mathrm{eff}}]\hat{\rho}_{R}=\hat{L}_{n,R,\mathrm{eff}}\hat{L}_{n,R,\mathrm{eff}}^{\dag}-\frac{1}{2}\{\hat{L}_{n,R,\mathrm{eff}}^{\dag}\hat{L}_{n,R,\mathrm{eff}},\hat{\rho}_{R}\}$ using Eq. (\ref{erS11}). Here we can negletct those fast oscillating cross terms. The long-range coherent hoppings in Eq. (\ref{SHeff}) are also can be neglected according to the same analysis. Then we define the hopping strength as
\begin{eqnarray}
J_{r(b)} = |j_{r(b)}|^{2}=\frac{\gamma|\Omega_{r(b)}|^{2}\eta_{r(b)}^{2}}{4\delta_{r(b)}^{2}+\gamma^{2}},
\end{eqnarray}
and obtain the effective Lindblad operators in the  Schr$\ddot{\text{o}}$dinger picture as 
\begin{align}
  L_{n, r} =&  \sqrt{n J_{r}} |g, n -1 \rangle\langle g, n |, \\
  L_{n, b} =&  \sqrt{n J_b} |g, n\rangle\langle g, n - 1|.
\end{align}
\bibliography{Citations}
\end{document}